\documentstyle[12pt,epsfig]{article}                                
\begin{document}
\titlepage                  
\vspace{1.0in}
\begin{bf}
\begin{center}

Nonlinear Oscillator  Hamiltonian from Nonlinear  Differential 
Equation and Calculation of Accurate Energy Levels

\end{center}
\end{bf}

\begin{center}
	      Biswanath Rath and P. Mallick

              Physics Department,North Orissa University ,Takatpur ,Baripada -757003, Odisha, INDIA

\end{center} 
\vspace{0.2in}

A new method for  generating analytical expression of quantum Hamiltonian  
from non-linear differential equation with stationary energy level  has been 
formulated.Further calculation of energy levels have been carried out 
analytically using and numerically using matrix diagonalisation method.

\vspace{0.2in}


\vspace{0.2in}

Key words- Non-lilear differential equation; energy level; matrix 
diagonalisation method,   Perturbation theory,Non-linear oscillator.

\vspace{0.2in}
\begin{bf}
I.Introduction
\end{bf}
\vspace{0.2cm}

In classical as well as quantum mechanics there are a only  few problems which
 can be solved exactly. It has been seen that many approximation methods have
 been applied to Harmonic oscillator with a view to bring new methods to 
limelight.However, to the best of our knowledge none of the published work [1-7]
 deals with the generation of quantum Hamiltonian from non-linear differentail equation.

In this communication, we present an innovative method to derive quantum Hamiltonian from non-linear differentail equation 

\begin{bf}
II.Method 
\end{bf}
\vspace{0.2cm}

In one dimension differential equation can be written as 

\[x^{..}+f(x)+ F(x^{.},x)=0    \hspace{1.0in}(1) \]

where $ f(x)$ is a function of $ x $ and $ F(x^{.},x)$ is a function of $x $ 
and it's  derivative with respect to time. The form of $f(x)$ and $F(x^{.},x)$
 can be considered as 
\[f(x)=\sum_{l}  \alpha_{l} x^{l}     \hspace{1.0in}(2) \]
and
\[ F(x^{.},x)=  \sum_{k,m} \lambda_{k,m} (x^{.})^{k} x^{m} \hspace{1.0in}(3) \]
The  equation of motion can be written as
\[\frac{d x^{.}}{dt} = -f(x) - F(x^{.},x)      \hspace{1.0in}(4) \]

Now using the relation 
\[\frac{d x^{.}}{dt} = -\frac{d V(x)}{dx}       \hspace{1.0in}(5) \]
and
\[H  = \frac{p^{2}} {2 } + V(x)       \hspace{1.0in}(6) \]
we can  get $ V(x) $ as
\[ V(x)= \int [f(x) + F(x^{.},x)] dx        \hspace{1.0in} \]
 In the above $H$ is considered as the Hamiltonian of the system in which $V(x)$ is the potential (potential energy) .
Now we can write  the Hamiltonian as  
   \[ H=\frac{ p^{2}}{2}+ \int [f(x)+ F(x^{.},x)] dx   \hspace{1.0in}(7) \]
In order to simplify the form of $H$, we use the formalism of second 
quantization and express $x$ and $p$ in terms of creation operator, $a^{+}$ 
and anihilation operator, $a$ as
   \[ x=\frac{1}{\sqrt 2w } [ a +a^{+}]     \hspace{1.0in}(8) \]
and
   \[ p=i {\sqrt\frac{w}{2}} [a^{+}-a]     \hspace{1.0in}(9) \]

Using equations (8) and (9), we can write Eqn. (7) as
 \[ H=\frac{ p^{2}}{2}+ \frac{1}{\sqrt 2w } \int [f(x)+ F(x^{.},x)] (da + da^{+} )  \hspace{1.0in}(10) \]
 
\begin{bf}
III.Hamiltonian in Second Quantization form  
\end{bf}
\vspace{0.2cm}
In the Eq(10) one has to use different forms of $f(x)$ and $F$ to get the
 Hamiltonian in second quantization form.

\begin{bf}
III.A.Linear $f(x)$   
\end{bf}

\vspace{0.2cm}

Considering  $f(x) = w^{2} x  $ and F($x^{.}$ x)= $\lambda (x^{.})^{2} x $,
the Hamiltonian in Eqn. (10) can be written as
\[  H =[2 a^{+} a+1]\frac{w}{2}+\frac{\lambda}{16}+\frac {\lambda}{16}[2(a^{+})^{2}a^{2} + 4 a^{+} a]-\frac {\lambda}{16}[(a^{+})^{4}+a^{4}] \hspace{1.0in}(12)\]
\begin{bf}
III.B.Non-Linear $f(x)$
\end{bf}

\vspace{0.2cm}

We also consider  $f(x) = w^{2} x + \lambda x^{3} $ and F($x^{.}$ x) =
$\lambda (x^{.})^{2} x $ and using Eqn. (10), the Hamiltonian (with $w$ = 1) can be written as
        \[ H(1) =\frac{[2 a^{+} a+1]}{2} +\frac{\lambda}{4}+\frac {\lambda}{2}  [(a^{+})^{2}a^{2} +2 a^{+}a] + \frac {\lambda}{8}[2(a^{+})^{3} a + 2a^{+} a^{3}+3a^{2}+3(a^{+})^{2}]     \hspace{1.0in}(13)\]

\begin{bf}
IV. Perturbation Theory
\end{bf}
 In this section, we use perturbation theory to calculate the energy levels up to second order. The Hamiltonian can be written interms of petrubed and unperturbed Hamiltonian as
   \[H = H^{'}_{0}  + H^{'} _{1}  \hspace{1.0in}(14) \]
where
\[  H_{0}^{'} =[2 a^{+} a+1]\frac{w}{2}+\frac{\lambda}{16}+\frac {\lambda}{16}[2(a^{+})^{2}a^{2} + 4 a^{+} a]     \hspace{1.0in}(15)\]
and
\[  H_{1}^{'} =- \frac {\lambda}{16}[(a^{+})^{4}+a^{4}]     \hspace{1.0in}(16)\]
Using standard pertuirbation theory[6,7], the expression for ground state energy
 upto second order  (considering $w=1$) is

\[  E_{0}=E_{0}^{(0)}+ E_{0}^{(1)}+E_{0}^{(2)}+..............\hspace{1.0in} (17) \]
 where
\[  E_{0}^{(0)}= \frac{ 1 }{2}+ \frac{\lambda }{16}     \hspace{1.0in} (18) \]
\[   E_{0}^{(1)}=0 \hspace{1.0in} (19) \]
\[ E_{0}^{(2)}=-\frac{3 \lambda ^{2}}  {32(4+2.5 \lambda)}  \hspace{1.0in} (20) \]

In Table-1 we present groundstate and first excited state energy of the
Hamiltonian in Eq(14) for $\lambda = 0.1$ .
Further if one considers the Hamiltonian of the form
   \[ H=\frac{ p^{2}}{2}+\int [w^{2} x+\lambda x^{3} + F(x^{.},x)] dx \hspace{1.0in} (21)\]
In this case  with $w=1$ we get
   \[H(1) = H^{'}_{0}  + H^{'} _{1}  \hspace{1.0in} \]
 where
\[  H_{0}^{'} =\frac{[2 a^{+} a+1]}{2} +\frac{\lambda}{4}+\frac {\lambda}{2}[(a^{+})^{2}a^{2} +2 a^{+}a]     \hspace{1.0in}(22)\]
and
\[  H_{1}^{'} = \frac {\lambda}{8}[2(a^{+})^{3} a + 2a^{+} a^{3}+3a^{2}+3(a^{+})^{2}]     \hspace{1.0in}(23)\]
In Table-1 we present groundstate and first excited state energy of the
Hamiltonian in Eq(14) for $\lambda = 0.1$ .

\begin{bf}
V. Matrix Diagonalization Method (MDM)
\end{bf}

In order to get accurate energy levels of this Hamiltonian (Eqn. (10)) we use
matrix diagonalisation method as follows. Now we solve the eigen-value relation [8-10]
\[ H\psi= E \psi  \hspace{2.0in} (24)\]
                 where \[ \psi= A_{m} |m>_{w}   \hspace{2.0in} (25)\]

For Hamiltonian in Eqn. (12), 

\[  H =[2 a^{+} a+1]\frac{w}{2}+\frac{\lambda}{16}+\frac {\lambda}{16}[2(a^{+})^{2}a^{2} + 4 a^{+} a]-\frac {\lambda}{16}[(a^{+})^{4}+a^{4}] \hspace{1.0in} \]

we obtain a three term recurrence relation [ as
     \[  A_{m+4} P_{m} + A_{m}R_{m}+ A_{m-4} T_{m}=0 \hspace{2.0in} (26)\]
where
     \[ P_{m}= <m|H|m+4>_{w}  \hspace{2.0in} (27)\]
     \[ R_{m}= <m|H-E|m>_{w}  \hspace{2.0in} (28)\]
     \[ T_{m}= <m|H|m-4>_{w}  \hspace{2.0in} (29)\]

Here $|m>_{w}$     is the mth state wave function of Harmonic oscillator with
parameter $w$. On solving the three term recurrence relation using matrix
diagonalisation method, we calculated first five energy levels for the
Hamiltonian given in Eqn. (12) and reflected the same in Table 1.

The Hamiltonian in Eqn. (13)(with $w$ = 1) can be written as
 \[ H(1) =\frac{[2 a^{+} a+1]}{2} +\frac{\lambda}{4}+\frac {\lambda}{2}  [(a^{+})^{2}a^{2} +2 a^{+}a] + \frac {\lambda}{8}[2(a^{+})^{3} a + 2a^{+} a^{3}+3a^{2}+3(a^{+})^{2}]     \hspace{1.0in} \]
In order to get the accurate energy eigen-values of this Hamiltonian, we use matrix diagonalisation method as stated above. In this case, we obtain a three term recurrence relation as
  \[ A_{m+2} Q_{m} + A_{m}R_{m}+ A_{m-2} S_{m}=0 \hspace{2.0in} (30)\]

where
     \[ Q_{m}= <m|H(1)|m+2>_{w}  \hspace{2.0in} (31)\]
     \[ R_{m}= <m|H(1)-E|m>_{w}  \hspace{2.0in} (32)\]
     \[ S_{m}= <m|H(1)|m-2>_{w}  \hspace{2.0in} (33)\]

\begin{bf}
VI. Results and Conclusion
\end{bf}

In the present paper, we accurately calculate the numerical values of energy
levels of the Hamiltonian obtained from non-linear differential equation using
matrix diagonalisation method. Further we noticed that the numerical value of
the ground state energies for the above two Hamiltonians nicely matches with theanalytical value. For example: the ground state energies of Hamiltonian
(Eqn.(6)) in the present method is found to be 0.506 029 which matches nicely
with the previously reported analytical value (0.506 03).
 Similarly, for the Hamiltonian in Eqn. (13),  the present convergent value
for the ground state is  0.523 76 whereas the previously reported analytical
value is 0.523 77. In our formalism, we practically developed a new method for generating Hamiltonian from any non-linear differential equation of second order.
Further one can calculate its enegy levels either numerically or analytically.

\pagebreak

\bf Table -I  \\
First two  energy levels of the Hamiltonian   Eq(12) and Eq(13) \ \\  

\begin{tabular}{|c|c|c|c|c|c|   }     \hline

n & $E_{n}^{(0)}$ &  $E_{n}^{(1)}$ &$E_{n}^{(2)}$ &$E_{n}$ &Hamiltonian  \\ \hline

0 &0.506 25  & 0    &-0.000 220    &0.506 03  & Eq(12) \\ 
1 &1.531 25  &0     &-0.001 077    &1.530 173 &    \\ 
  &            &      &         &           &     \\ \hline  
0 & 0.525   &0     &-0.001 222  &0.523 778  & Eq(13)  \\ 
1 &1.625   & 0    &-0.009 375   &1.615 625  &    \\ \hline

\end{tabular}

\bf Table -II  \\
First five  energy levels of the Hamiltonian   (Eq(12))  \ \\

\begin{tabular}{|c|c|c|c|c|c|   }     \hline

n & Size(9x9) & Size(19x19)& Size(29x29) & Size(39x39)   \\ \hline

0 &0.506 029 039&0.506 029 038&0.506 029 038 &0.506 029 038  \\
1 &1.530 172 680&1.530 169 441&1.530 169 441 &1.530 169 441   \\
2 &2.578 092 128&2.578 076 954&2.578 076 954 &2.578 076 954   \\
3 &3.649 049 855&3.648 997 679&3.648 997 673 &3.648 997 673  \\
4 &4.742 400 314&4.742 253 427&4.742 253 401 &4.742 253 401   \\ \hline

\end{tabular}

\bf Table -III  \\
First five  energy levels of the Hamiltonian   (Eq(13))  \ \\

\begin{tabular}{|c|c|c|c|c|c|   }     \hline

n & Size(9x9) & Size(19x19)& Size(29x29) & Size(39x39)   \\ \hline

0 &0.523 767 849&0.523 767 849&0.523 767 849 &0.523 767 849  \\
1 &1.615 478 755&1.615 478 611&1.615 478 611 &1.615 478 611   \\
2 &2.791 344 321&2.791 342 192&2.791 342 192 &2.791 342 192   \\
3 &4.044 414 426&4.044 094 127&4.044 094 126 &4.044 094 126  \\
4 &5.369 741 578&5.368 297 477&5.368 297 470 &5.368 297 469   \\ \hline

\end{tabular}

\pagebreak


\begin{thebibliography}{99}
\bibitem{ Ginsburg} C.A. Ginsburg,  Phys. Rev. Lett {\bf 48} (1982) 839.
\bibitem{Bender} C.M. Bender and T.T. Wu, Phys. Rev. {\bf 184} (1965) 1234.
\bibitem{ Weniger } E.J. Weniger, Phys. Rev. Lett. {\bf 77} (1996) 2862.
\bibitem{ Janke  } W. Janke and H. Kleinert, Phys. Rev. Lett.  {\bf 75 } (1995) 2787 (1995)
\bibitem{Schiff }L.I. Schiff, Quantum Mechanics, {\bf 3rd ed}(McGraw Hill, singapore, 1985).
\bibitem{Rath } B.Rath,  Phys. Rev. {\bf A42(5)} (1990) 2520.
\bibitem{Rath } B.Rath,  Eur. J. Phys. {\bf 11 } (1990) 184.
\bibitem{Rath } B.Rath, Int. J. Mod. Phys. {\bf A 14(13)} (1999) 2103.
\bibitem{Rath } B.Rath,  J. Phys. Soc. Jpn. {\bf 67(9)} (1998) 3044.
\bibitem{Rath } B.Rath,  O. J. Phys. {\bf 19(2)} (2012) 157.



\end{thebibliography}
\end{document}